\documentclass[a4paper,fleqn,usenatbib]{mnras}

\usepackage{mathptmx}

\usepackage[T1]{fontenc}
\usepackage{ae,aecompl}

\usepackage[percent]{overpic}
\usepackage{graphicx}	
\usepackage{amsmath}	
\usepackage{amssymb}	
\usepackage{pdflscape}

\title[DART tail]{Ground-based observability of Dimorphos 
  DART impact ejecta: Photometric predictions}  

\author[F. Moreno]{
  Fernando Moreno,$^{1}$\thanks{E-mail: fernando@iaa.es}
  Adriano Campo Bagatin$^{2}$, Gonzalo Tancredi$^{3}$,
\newauthor
  Po-Yen
  Liu$^{2}$, Bruno Dom\'\i nguez$^{3}$
  \\ 
$^{1}$Instituto de Astrof\'\i sica de Andaluc\'\i a, CSIC, Glorieta de
  la Astronom\'\i a s/n, 18008 Granada, Spain \\
$^{2}$Departamento de F\'\i sica, Ingenier\'\i a de Sistemas y Teor\'\i
  a de la
  Se\~nal, Universidad de Alicante, San Vicent del Raspeig, 03690 
  Alicante, Spain\\
$^3$Departamento de Astronom\'\i a, Facultad de Ciencias, Igu\'a 4225,
  11400 Montevideo, Uruguay
}

\pubyear{2021}

\begin{document}
\label{firstpage}
\pagerange{\pageref{firstpage}--\pageref{lastpage}}
\maketitle

\begin{abstract}

The Double Asteroid Redirection Test (DART) is a NASA mission intended
to crash a projectile on Dimorphos, the secondary component of the
binary (65803) Didymos system, to study its orbit deflection. As a
consequence of the 
impact, a dust cloud will be   
be ejected from the body, potentially forming a transient coma- or
comet-like tail 
on the hours or days 
following the impact, which might be observed using ground-based
instrumentation. Based on the mass and speed of the impactor,
and using known scaling laws, the total mass ejected can be
roughly estimated. Then, with the aim to provide approximate expected
brightness  
levels of the coma and tail extent 
and morphology, we have propagated  
the orbits of the particles ejected by integrating their equation of
motion, and have used a Monte Carlo approach
to study the evolution of the coma and tail brightness. For  
typical power-law particle size distribution of index --3.5, with
radii r$_{rmin}$=1
$\mu$m and r$_{max}$=1 cm,   
and ejection speeds near 10 times the escape velocity of 
Dimorphos, we predict an increase of brightness of $\sim$3 magnitudes
right after the impact, and a decay to pre-impact
levels some 10 days after. That would be the case if the prevailing 
ejection mechanism comes from the impact-induced seismic
wave. However, if most of the ejecta is released at speeds of the
order of $\gtrsim$100 $\mathrm{m\; s^{-1}}$, the observability of the event
would reduce to a very short time span, of the order of one day or shorter.

\end{abstract}




\section{Introduction}

The Double Asteroid Redirection Test (DART)
\citep{2018P&SS..157..104C} is a planetary defence 
mission launched November 24th, 2021 by NASA. This mission is aimed to
test asteroid-deflecting technology by crashing a $\sim$535 kg projectile on
the surface of Dimorphos, a $\sim$160 m diameter body orbiting the
larger (780 m) main-belt asteroid (65803) Didymos. During the first
few minutes after the impact, the Light Italian CubeSat for Imaging
Asteroids, or LICIACube \citep{2021P&SS..19905185D}, will be
taking images of the early ejecta. At the planned collision
time (2022 September 26th, 23:14 UT) and the few days following the
impact, this binary system will be well placed for southern hemisphere
observers, and at fairly high galactic latitudes (--50$^\circ$ to
--30$^\circ$ between 2022 September 26, and October 8), so that 
relatively clear background star fields are foreseen. Ground-based
observations of the dust cloud generated after the impact and its
evolution with time will provide information on the amount of mass
released, the ejection speeds, and the physical properties of the dust
particles. In order to estimate some of these parameters in
advance, which might aid to plan the observations, we have performed
simulations of 
the evolution of the dust coma and tail that will appear after the impact, its
brightness level, extent, and morphology. In section 2 we describe the
physical model used, and the governing equation of motion of the
ejected particles.  In section 3 we
describe the method used to integrate the equation of motion, and a
comparison with an independent N-body simulator is performed for
validation purposes. Section 4 gives a description of the procedure to build
up the dust tail synthetic images, along with the particle physical
properties adopted and the expected dust mass released. In Section 5,
the dust tail
code is further validated, for sufficiently high ejection speeds,
with our classical Monte Carlo dust tail 
code that we routinely use to fit cometary and active asteroid tail 
images \citep[see e.g.][and references
    therein]{2021MNRAS.506.1733M}. Section 6 shows the results
obtained concerning dust 
tail brightness simulations as well as synthetic lightcurves as a
function of the ejecta speed. Finally, the conclusions of this work
are given in Section 7. 

\section{Physical model of the binary system}

The DART impact is foreseen to occur  
on 2022 September 26th, 23:14 UT (JD=2459849.46806). The impactor mass
is assumed at $m_i$=535 kg, and its 
speed at impact time, $v_i$=6.7 $\mathrm{km\; s^{-1}}$. We assume
that the binary components 
of the asteroid system are both spherical, and adopt physical
parameters from the latest ESA's Hera Mission
\citep{2018AdSpR..62.2261M} Didymos 
Reference Model available (12/03/2020, 
 see Table ~\ref{PhysPar}). The 
binary system has a mutual orbit 
period of $P_{sec}$=11.9217 h and a semi-major axis of $a_{sec}$=1.19
km. Hence, using
the Kepler's third law, the total system mass becomes 
$M_T$=5.4$\times$10$^{11}$ kg. The Dimorphos mass, assuming the same
density as Didymos, will be $M_{sec}$=5$\times$10$^9$  
kg. The Didymos rotation is retrograde, the north 
pole of the spin axis pointing to ecliptic coordinates 
$(\lambda_e,\beta_e)$=(320.6$^\circ$,--78.6$^\circ$). This orientation
correspond to the latest orbit solution 
JPL 104 for Dimorphos \citep{NaiduJPL}, whose orbit is contained in the Didymos
equatorial plane. The heliocentric ecliptic coordinates and velocity
of Didymos at the 
impact time are taken from the JPL Horizons on-line Ephemeris System\footnote{
https://ssd.jpl.nasa.gov}. The 
Dimorphos coordinates and velocity components relative to Didymos
body centre at impact time are also taken from the JPL Horizons web
site. For completeness, these coordinates and velocities
are shown in Table ~\ref{coordinates}. 
  \begin{table}
	\centering
	\caption{Physical parameters for the Didymos-Dimorphos
          system, adopted from latest Hera Didymos 
          Reference Model available (12/03/2020).}
        \label{PhysPar}
	\begin{tabular}{ll}
          Parameter & Value \\
          \hline   
          Didymos diameter   & 780$\pm$30 m  \\
          Dimorphos diameter & 164$\pm$18 m  \\
          Semimajor axis of Dimorphos orbit    & 1.19$\pm$0.03 km  \\
          Dimorphos orbital  period & 11.9217$\pm$0.0002 h  \\
          Didymos density    & 2170$\pm$350  $\mathrm{kg\; m^{-3}}$ \\
          \hline
	\end{tabular}
   \end{table}

  \begin{table*}
	\centering
	\caption{Heliocentric ecliptic coordinates and velocity at impact time
          (JD=2459849.46806). The Dimorphos data are relative to
          Didymos body centre. All coordinates are taken from JLP
          Horizons on-line Ephemeris System. Units are  $\mathrm{au}$
          for position and  $\mathrm{au\; day^{-1}}$ for velocity components.}
        \label{coordinates}
	\begin{tabular}{lll}
          \multicolumn{3}{c}{Didymos}\\
          \hline
   X=1.040512227512945E+00 & Y=  9.017812950063080E-02&
   Z=-5.774272064726729E-02 \\
        $V_x$=-4.229184367475531E-03 & $V_y$=1.917344670139128E-02  &
   $V_z$=5.728269981716490E-04\\ 
\hline
        \multicolumn{3}{c}{Dimorphos}\\
	  \hline
      $\Delta$X=--5.604615661026744E-09 & $\Delta$Y=-5.736939008292839E-09&
      $\Delta$Z=-1.390204184153139E-10\\
     $\Delta V_x$=-7.091500670770579E-08  &   $\Delta
          V_y$=6.976345650818189E-08&   $\Delta  V_z$=-1.997791163791703E-08 \\
		\hline
	\end{tabular}
\end{table*}

The heliocentric position of Didymos at any time during the particle
orbital integration is computed from its orbital elements. For
Dimorphos, we compute its position relative to Didymos
from its orbital elements related to Didymos body centre. The orbit
of Dimorphos relative to Didymos becomes elliptical, with a small 
eccentricity of $\epsilon$=0.04, in line with previous estimates \citep[see
  e.g.][]{2009Icar..200..531S,2012AJ....143...24F,2020Icar..34813777N}. 

For the particle orbit integration, we use the heliocentric ecliptic
coordinate system relative to Didymos body centre, denoted by
lowercase letters $(x,y,z)$. We define a 
colatitude-longitude $(\theta,\phi)$ grid on Dimorphos surface so that
the initial 
position of the particle on Dimorphos is given by: 
$x_i=R_{sec}\cos \phi \sin \theta$, $y_i=R_{sec}\sin \phi \sin\theta$, and
$z_i=R_{sec}\cos \theta$, where $R_{sec}=D_{sec}/2$. The initial
coordinates of the particle relative to Didymos are given by:

\begin{equation}
  \begin{aligned}
    x_0=x_i+\Delta X \\  y_0=y_i+ \Delta Y \\  z_0=z_i+ \Delta Z \\ 
\end{aligned}
\label{eqinit}
\end{equation}

\noindent
where ($\Delta$X,$\Delta$Y,$\Delta$Z) are the initial coordinates of
Dimorphos relative to Didymos (see Table ~\ref{coordinates}). The
heliocentric ecliptic coordinates of the particle are given by:

\begin{equation}
  \begin{aligned}
X_p=x+X_D\\Y_p=y+Y_D\\Z_p=z+Z_D\\ 
\end{aligned}
\label{eqinit}
\end{equation}

\noindent
where $(X_D,Y_D,Z_D)$ are the heliocentric coordinates of Didymos.

Another Didymos-centred reference frame is defined as the one having
$(\xi,\eta,\zeta)$ axes, where the $\xi$ axis points to the same
direction and sense as the Sun-to-Didymos radius vector, $\eta$ is
contained in 
the asteroid orbital plane and is directed opposite to its  
orbital motion, and  $\zeta$ is perpendicular to the orbital
plane and chosen to form a right-handed set with $\xi$ and $\eta$
axes \citep[see][]{1968ApJ...154..327F}. This reference frame is
useful to build up the dust tail as seen from Earth, by 
projecting the particle $(\xi,\eta,\zeta)$ coordinates onto the
so-called $(N,M)$ photographic plane. The equations giving such
projection are given in \cite{1968ApJ...154..327F}. This
transformation also involves 
the computation of the Earth coordinates at the time of observation
$(\xi_E,\eta_E,\zeta_E)$, 
which are computed from its heliocentric ecliptic coordinates,
available in the JPL Horizons on-line Ephemeris System. The 
transformation of heliocentric ecliptic coordinates to the 
$(\xi,\eta,\zeta)$ asteroid-centred reference frame is performed
by standard methods.

In addition to the gravity forces from the bodies involved, the only
non-gravitational force acting on a particle that we consider in the
model is the solar radiation pressure. Thus, the Poynting-Robertson
drag on the particles is only important in very long-term 
dynamics, and is therefore neglected. The Lorentz acceleration, that
would act on the smallest grains of the assumed size distribution, can
be written as \citep{2019Icar..319..540P}: 

\begin{equation}
a_L=\frac{12 \epsilon_0 V_P}{C^2} \beta^2 \frac{\rho_p}{Q_{pr}^2} B_{\phi} V_{sw}
\label{eqLorentz}
\end{equation}

\noindent
where $\epsilon_0$ is the vacuum permittivity, $V_P$ is the potential
on the grain surface, assumed here as $V_P$=+5 V
\citep{1981A&A....99....1M, 1994A&A...286..915G},
$C$=5.76$\times$10$^{-4}$ $\mathrm{kg\; m^{-2}}$, $\rho_p$ is the
particle density, assumed at having the same value as the bulk density
of the asteroid, $Q_{pr}$ is the
scattering efficiency for radiation pressure, which takes the value
$Q_{pr}$=1, $B_{\phi}$ is the
azimuthal component of the interplanetary magnetic field, which has a
mean value of $B_{\phi}$=3 nT at 1 au from the Sun
\citep{2000JGR...10510303L}, and $V_{sw}$ is the solar wind speed, which
is taken as 400 $\mathrm{km\; s^{-1}}$ 
\citep{1994A&A...286..915G}. The quantity $\beta$ is defined as the
ratio of radiation pressure force to gravity force, as  
$\beta=F_{rad}/F_{grav}=C_{pr}Q_{pr}/(2\rho_p a_p)$, where
$C_{pr}$=1.19$\times$ 10$^{-3}$ kg m$^{-2}$, and $a_p$ is the grain
radius.  For the Lorentz 
acceleration, we then obtain $a_L=0.004 \beta^2$ $\mathrm{m\;
  s^{-2}}$. The radiation pressure 
force is given by $a_{RP}=(1-\beta)G M_\odot/r_h^2$, where $G$ is the
gravitational constant, $M_\odot$ is the Sun mass, and $r_h$ is the
heliocentric distance of the grain. At $r_h$= 1 au, we have
$a_{RP}=0.006(1-\beta)$ $\mathrm{m\;s^{-2}}$.  Therefore, at the lower
limit of the size distribution ($a_p$=1 $\mu$m, see section 4), $a_{RP}$
is higher than $a_L$ in more than an order of magnitude. The ratio
$a_{RP}/a_L$ increases as $\beta$ decreases, or, equivalently, $a_p$
increases. We then neglect the Lorentz force in our approach. Hence,
the equation governing the trajectory of each individual grain can be 
written as:

\begin{equation}
\begin{aligned}  
\frac{d^2\mathbf{r_d}}{dt^2} = & 
W_1\frac{\mathbf{r_d}}{r_d^3}
+W_2\frac{\mathbf{r_d}-\mathbf{r_s}}{\|\mathbf{r_d}-\mathbf{r_s}\|^3}
+ W_3\left[ \frac{\mathbf{r_s} - \mathbf{r_d}}{\|\mathbf{r_s}
 - \mathbf{r_d}\|^3} - \frac{\mathbf{r_s}}{r_s^3}\right] + \\
  &  W_4 \left[ \frac{\mathbf{r_{dsec}}}{r_{dsec}^3} -
    \frac{\mathbf{r_{sec}}}{r_{sec}^3} \right]  
\end{aligned}
\label{eqtraj}
\end{equation}

\noindent
where $\mathbf{r_d}$ is the Didymos-to-dust
grain vector, $\mathbf{r_s}$ is the Didymos-to-Sun vector, 
$\mathbf{r_{dsec}}$ is the vector from the dust grain to Dimorphos,
and $\mathbf{r_{sec}}$ is the Didymos-to-Dimorphos
  vector. We have used the fact that
  $\mathbf{r_s}$=$\mathbf{r_d}$ + $\mathbf{r_{ds}}$, where
  $\mathbf{r_{ds}}$ is the vector from the dust grain to the
  Sun. Figure ~\ref{geometry} provides a schematic drawing of the
  vectors used.  The other terms are $W_1=-GM_P$, $W_3=GM_\odot$
  ($M_P$ is Didymos mass), and $W_4=GM_{sec}$. The remaining term, $W_2$,
  is given by: 

\begin{figure}
     \includegraphics[clip, trim=1.5cm 6.6cm 2.5cm 2cm,width=0.6\textwidth]{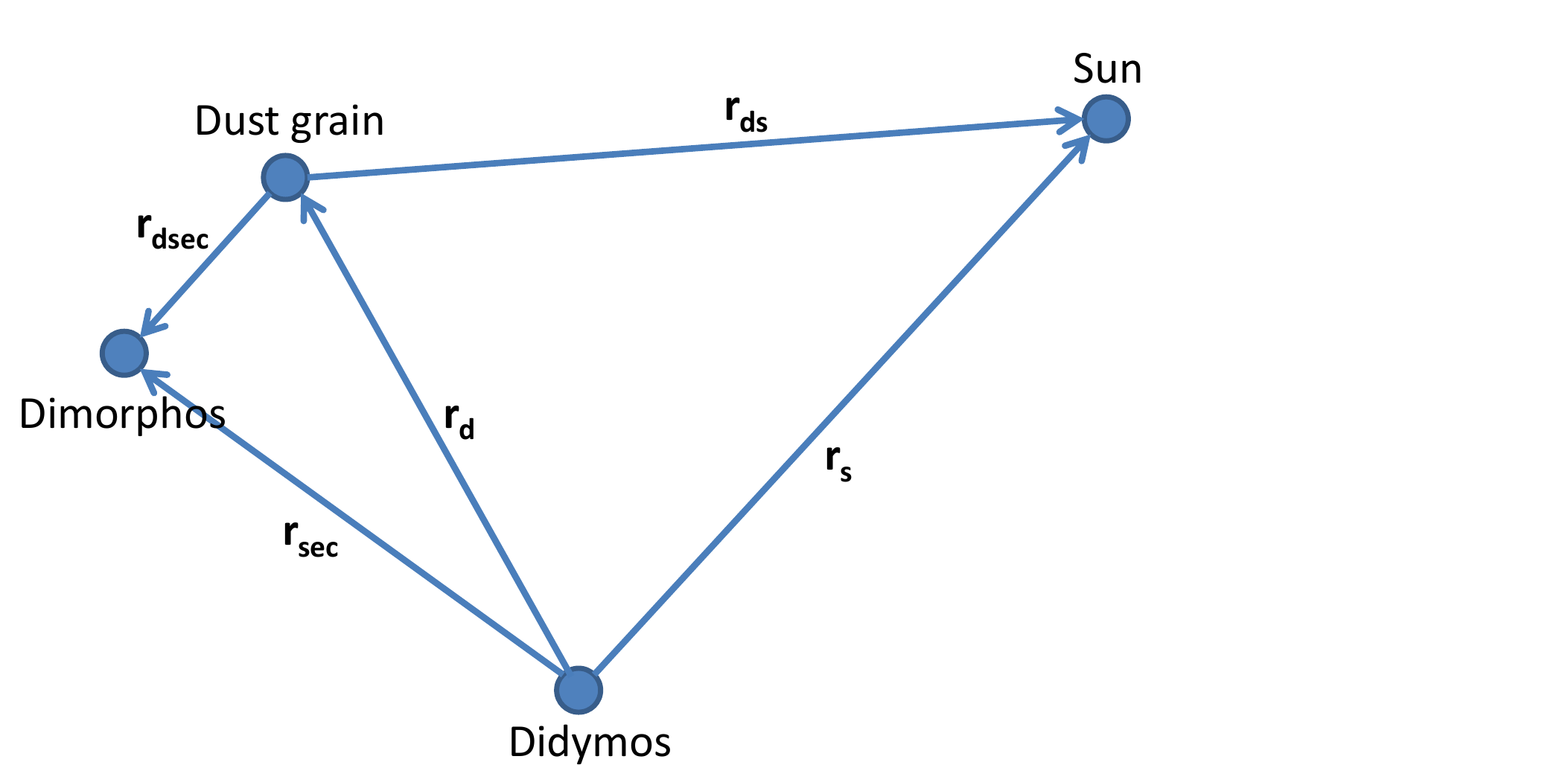}
     \caption{Schematic drawing of the vectors shown in Equation
       ~\ref{eqtraj}.}
    \label{geometry}
\end{figure}

\begin{equation}
  W_2 = \frac{Q_{pr}}{c} \frac{E_s}{4\pi} \frac{\pi d^2}{4 m_p}
\label{W2}
\end{equation}

\noindent
In equation ~\ref{W2}, $c$ is the speed of light,
$E_s$=3.93$\times$10$^{26}$ W is 
the total power radiated by the Sun, $d$ is the
particle diameter ($d$=2$a_p$),
and $m_p$ is the particle mass ($m_p=\rho_p (4/3)\pi a_p^3$).

\section{Orbital simulations}

\begin{figure*}
  \begin{tabular}{cc}
    \begin{overpic}[angle=-90,width=0.4\textwidth]{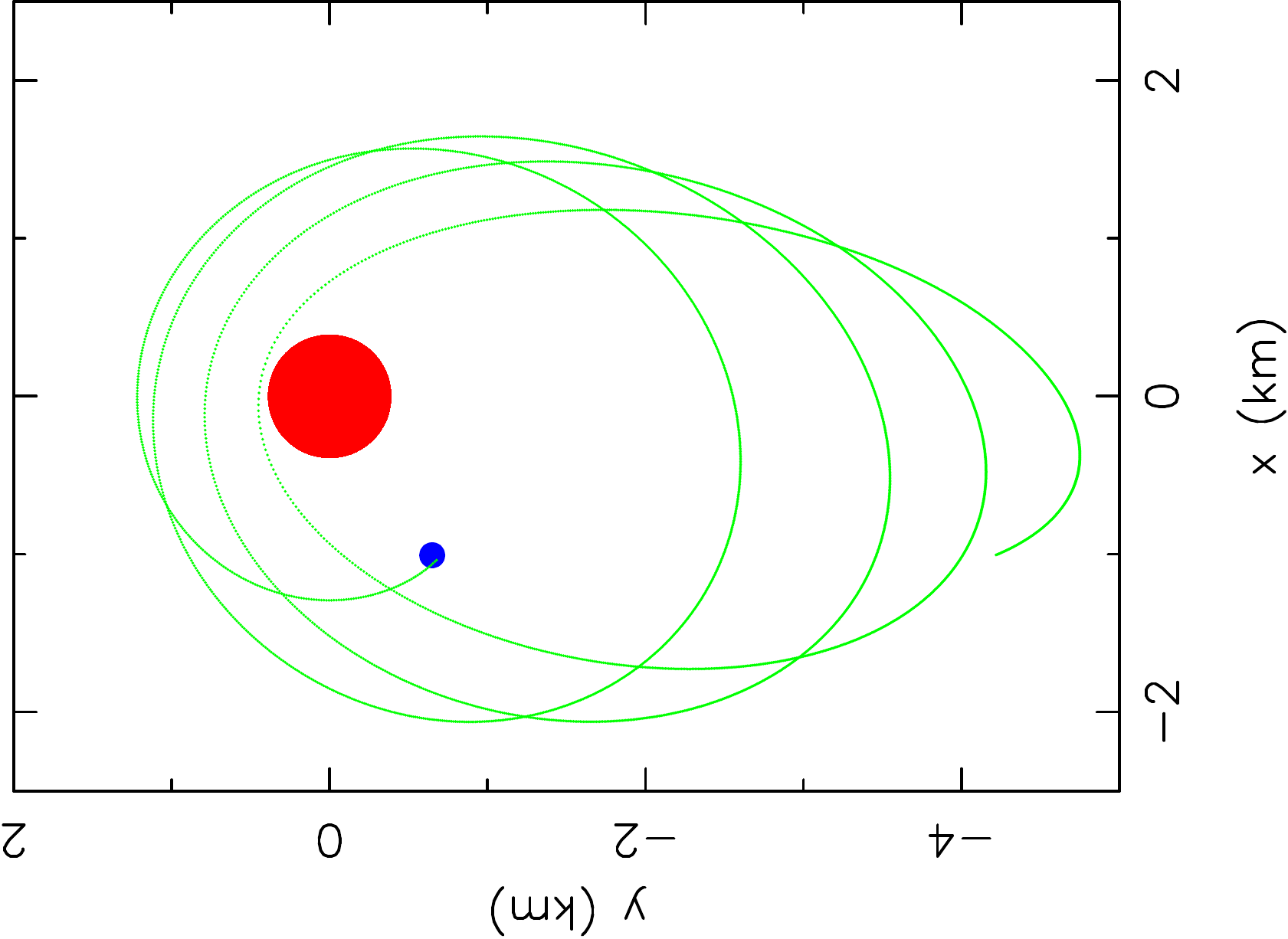}
      \put (67,89) {\Large (a)} 
    \end{overpic} &    
    \begin{overpic}[angle=-90,width=0.4\textwidth]{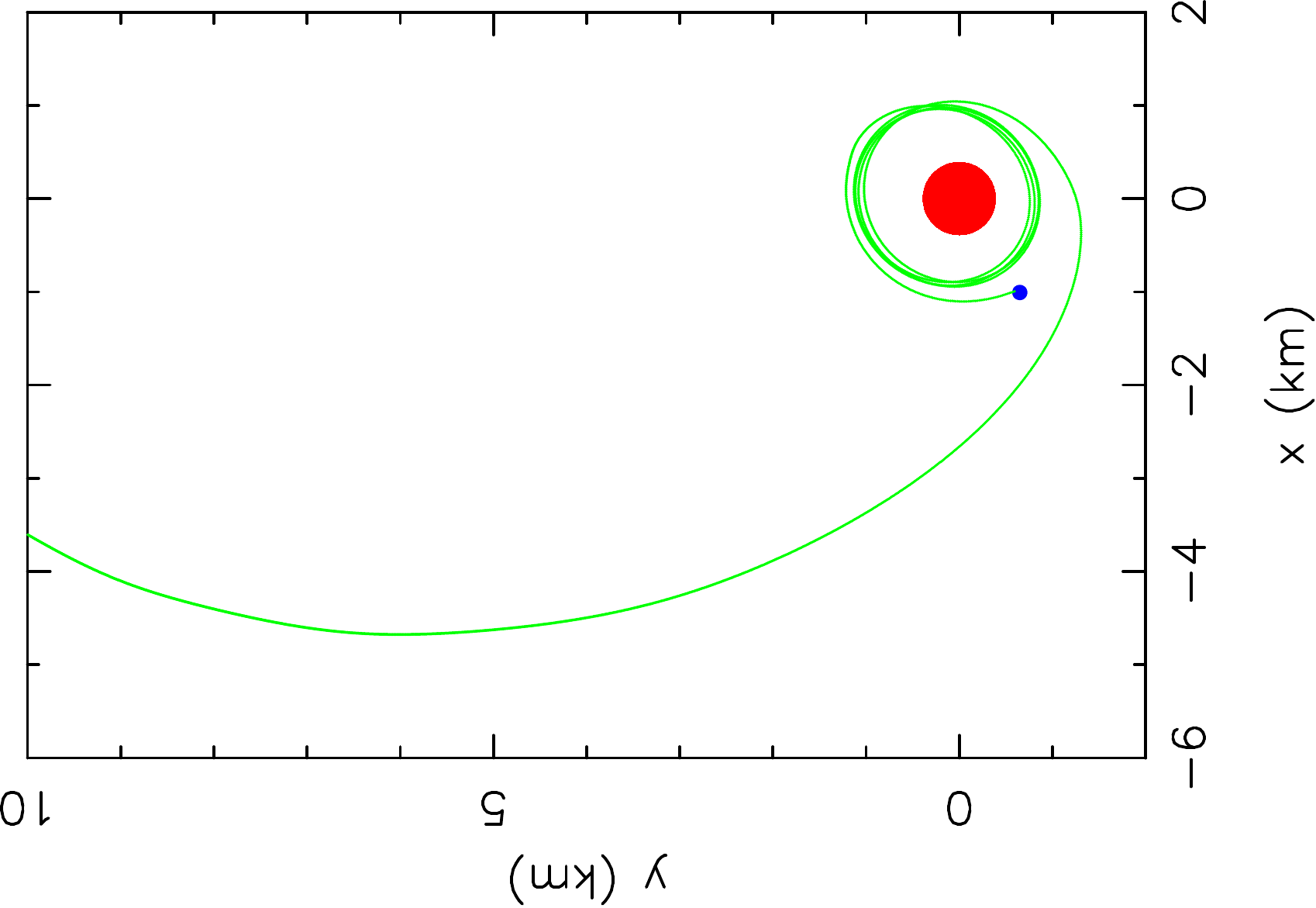}
      \put (61,89) {\Large (b)} 
\end{overpic} \\
    \begin{overpic}[angle=-90,width=0.5\textwidth]{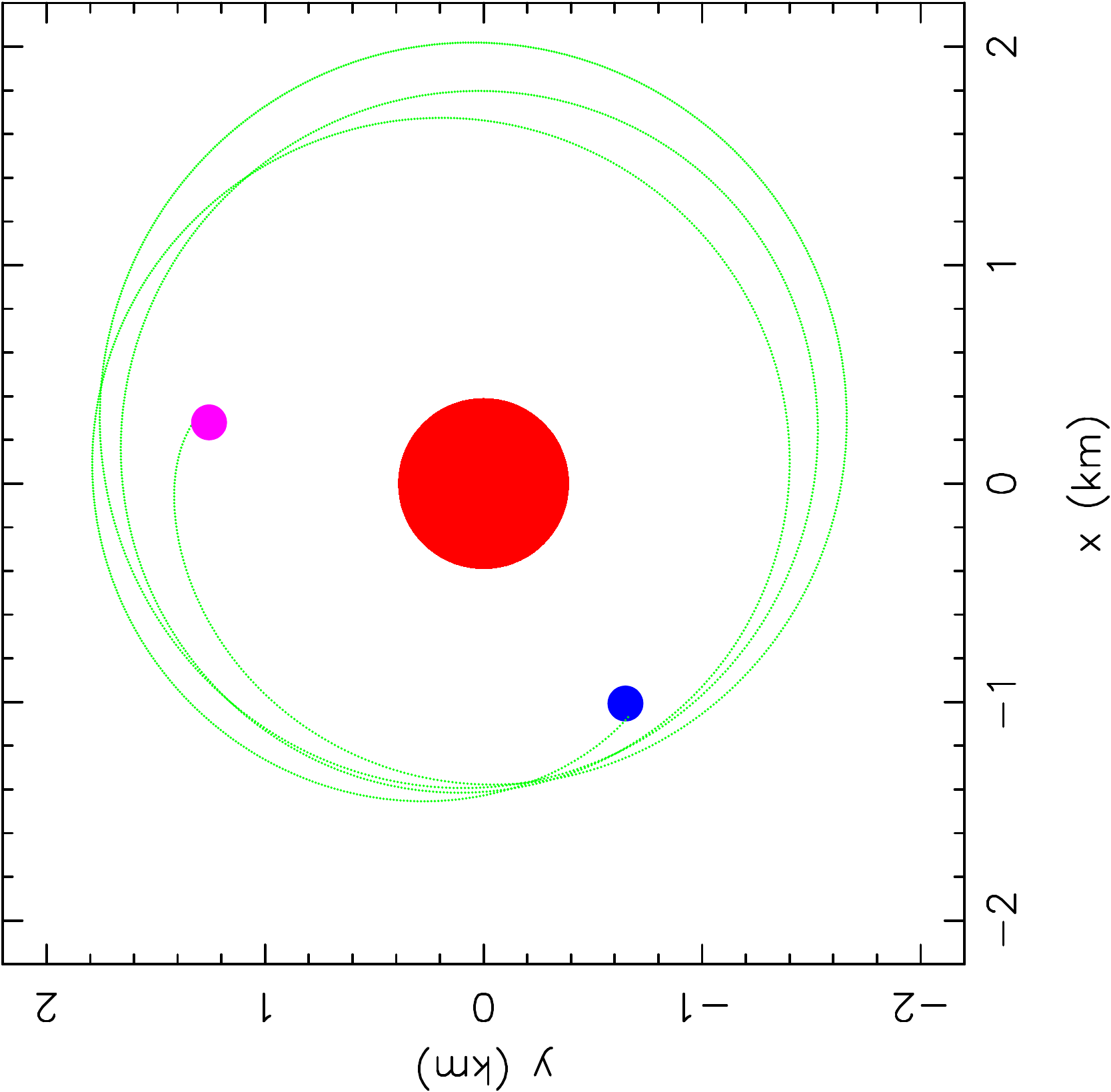}
      \put (90,85) {\Large (c)} 
    \end{overpic} &    
    \begin{overpic}[angle=-90,width=0.5\textwidth]{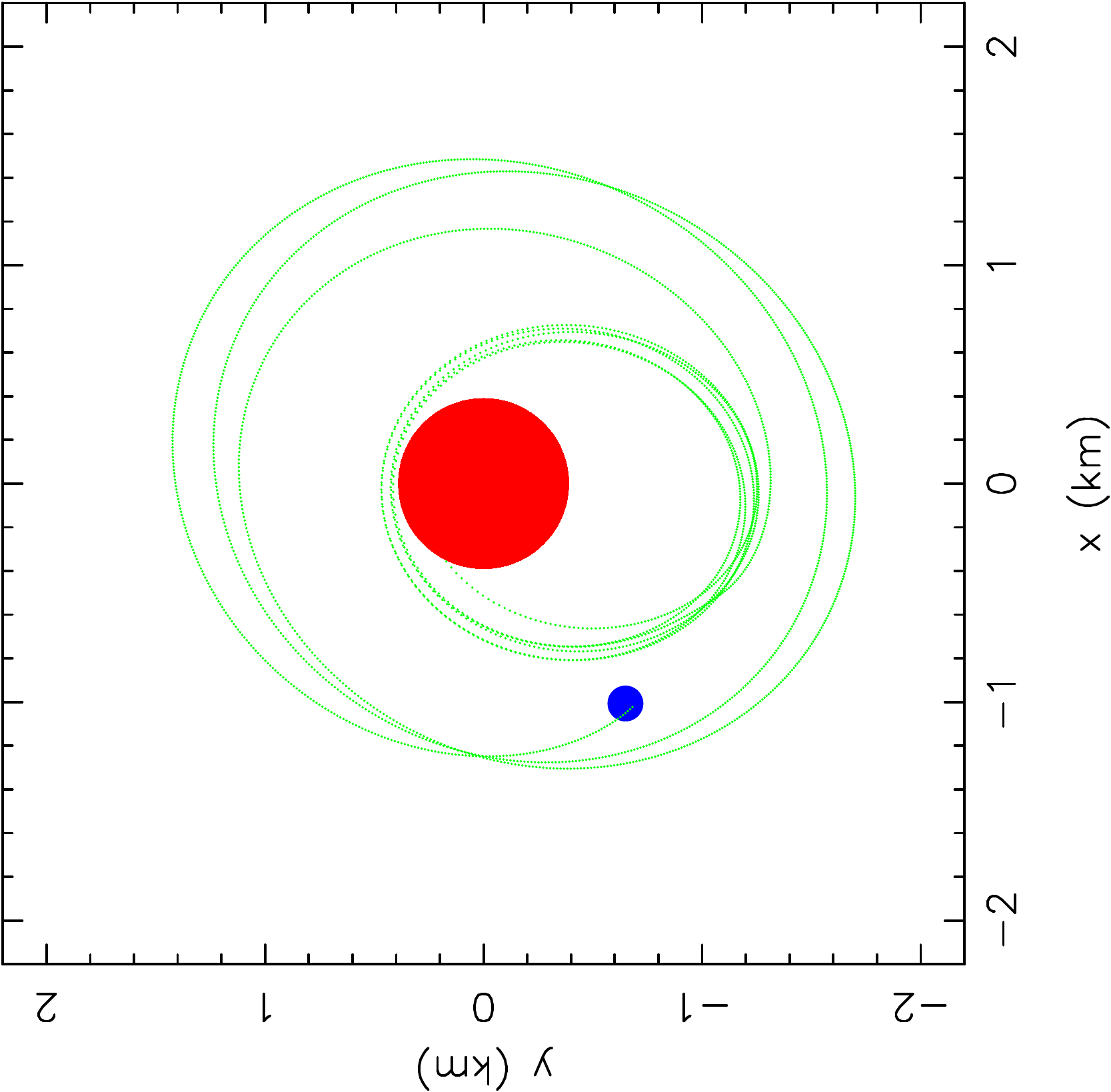}
      \put (90,85) {\Large (d)}
    \end{overpic} \\
    \end{tabular}
    
  \caption{Sample orbits of $r$=1 cm particles ejected at Dimorphos escape
  velocity from various locations on 
  Dimorphos surface, projected to the $x-y$ plane, and 
  integrated from impact time to October 2nd, 2022.  The initial positions of
  Didymos (large red filled circle) and 
  Dimorphos (small blue filled circle) are given. The orbits are drawn 
  as green lines.  (a) Orbit of a particle which is
  orbiting the binary asteroid and remains close to the system at the
  end of the integration. (b) A particle that leaves the system,
  contributing to the far tail region brightness. (c) A particle colliding with
  Dimorphos before the end of integration. The position of Dimorphos
  at the collision time is drawn as a purple circle. (d) A particle
  that collides with Didymos before the end of integration.} 
    \label{orbits}
\end{figure*}

To compute the trajectory of each particle, equation ~\ref{eqtraj} is
integrated numerically following a fourth-order 
Runge-Kutta ({\texttt{RK4}}) method. A constant time step of 100 s was
found appropriate 
to conduct the simulations. At each time step, we check whether the
particle is within the shadowed region of Didymos, so that the
radiation pressure was set to zero in that region. Since the geometrical cross
section of Dimorphos is very small, and in order to save computational
time, we do not take into account its projected shadow in the
calculations. The fate of each particle can 
be either (1) a collision with Didymos, (2) a collision with
Dimorphos, or (3) released to outer space, feeding up the dust
coma/tail. Sample trajectories are shown in Figure
~\ref{orbits}. Small 
particles tend to leave the binary system very fast, feeding up the tail,
while large particles stay orbiting the system for some time, or
eventually collide either with Didymos or with Dimorphos.  

\begin{figure*}
     \includegraphics[angle=-90,width=\textwidth]{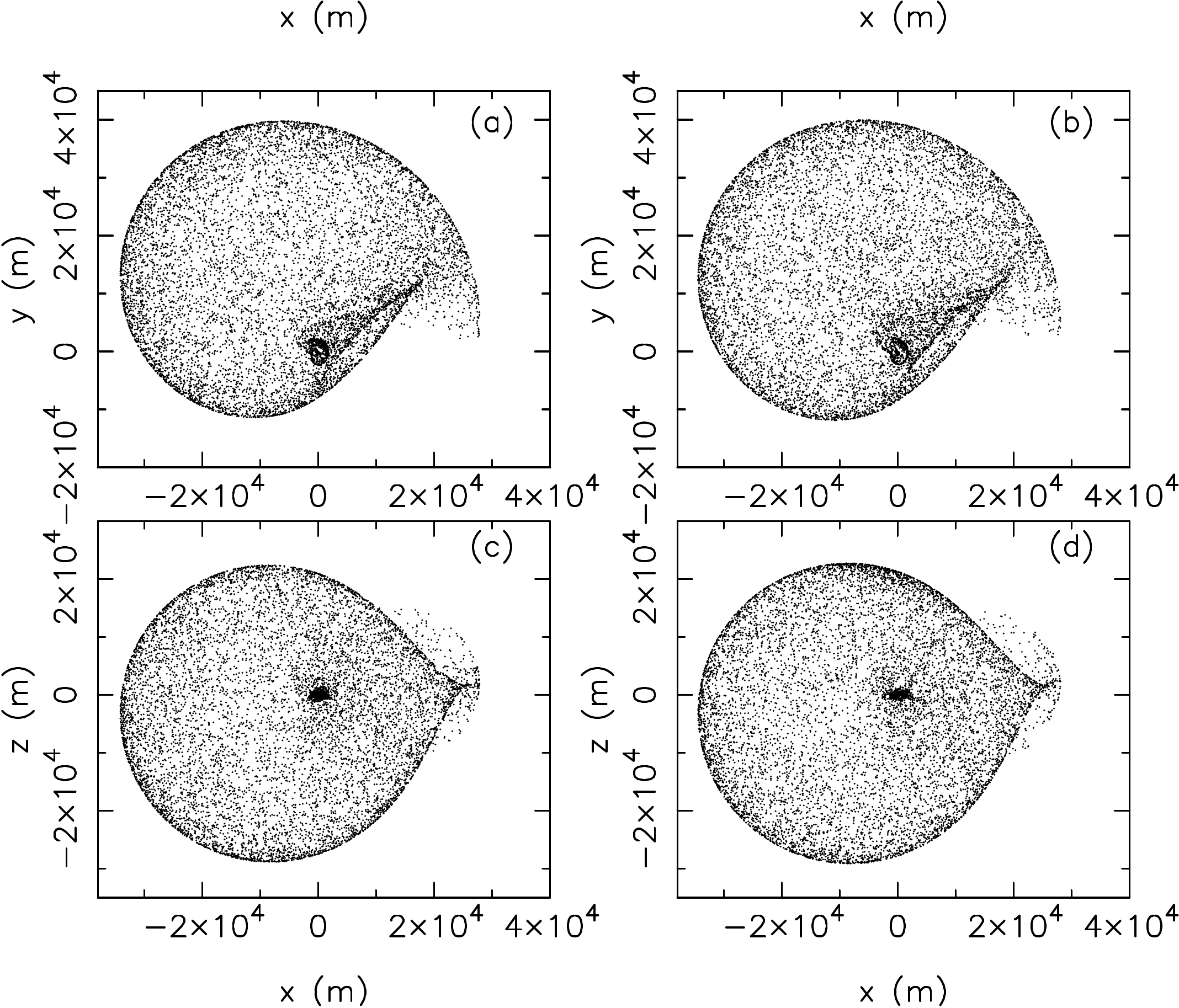}
     \caption{Final positions, projected on the 
       x-y planes (panels (a) and (b)) and x-z planes (panels (c) and
       (d)), relative to Didymos position, on September 28th, 2022, 
       of 10$^4$ particle of $r$=1 cm  
       ejected randomly in all directions from Dimorphos at the impact
       time. The ejection speed is four 
       times the escape velocity of Dimorphos, i.e., $v$=0.36
       $\mathrm{m\; s^{-1}}$. Panels (a) and (c) are the results 
       from the {\texttt{RK4}} code, and panels (b) and (d) are the
       corresponding calculations using 
       {\texttt{MERCURY}} 
       \citep{1999MNRAS.304..793C} code.} 
    \label{validation}
\end{figure*}

With the purpose of validating 
the results obtained, we have also performed test simulations using the
{\texttt{MERCURY}} N-body software package for orbital dynamics
\citep{1999MNRAS.304..793C}.  In addition to Didymos, Dimorphos, and
the corresponding particle, the planets Mercury, Venus, the Earth-Moon
system, Mars, and 
Jupiter were included in the simulations. While the effect of all
those planets in the short time integration interval of the ejecta
that we provide is
minimal, we 
keep the procedure as it is for future longer-term studies.
The radiation pressure force was set
by writing a dedicated user-defined force routine in the computer
code. A  Bulirsch-St\"oer 
integrator was used, with a time step of 10$^{-3}$ days (86.4 s), i.e., 
similar to the 100 s used for the {\texttt{RK4}} simulations above. The initial
heliocentric ecliptic coordinates and velocities of all the bodies 
were taken from the JPL Horizons on-line Ephemeris System as described
above. The initial heliocentric 
position and velocity components of Dimorphos and the particle
were also set as indicated above, assuming randomly selected radial direction
ejection from Dimorphos spherical surface. Figure ~\ref{validation}
shows a comparison of the fate of 1 cm particles, on September 28th, 2022,
ejected at $v$=0.36  $\mathrm{m\; s^{-1}}$ from Dimorphos surface 
using both the {\texttt{RK4}} and {\texttt{MERCURY}} procedures, where
an excellent agreement can be seen. 

\section{Dust tail build-up}

The last step at the end of the orbital integration is the calculation
is the projection of the particle 
position onto the sky $(N,M)$ plane, as described  previously. The
$(N,M)$ coordinates are finally rotated to equatorial 
coordinates $(RA,DEC)$ through the asteroid position angle, so that
the tail images are provided in the conventional North up, East to
left orientation. In this way, they can be directly compared with 
Earth-based telescope observations. 

The brightness of the tail is computed by adding up the
contribution of each particle of the large set of particles that are
released from Dimorphos surface, taking into account the particle
physical properties, the size
distribution, and the total dust mass ejected. This procedure has been
outlined in many previous papers \citep[see e.g.][and references
  therein]{2021MNRAS.506.1733M}. For completeness, however, a
    description of the method is given in the Appendix.

We assume that the physical
properties of the particles are in line with the 
  Didymos known properties. A geometric albedo of
  $p_R$=0.16$\pm$0.04 has been estimated for 
  Didymos \citep {2006Icar..181...63P}.  Then, a linear phase
  coefficient of $\phi=0.013-0.0104 \ln p_R$=0.032  $\mathrm{mag\; deg^{-1}}$
   is
  calculated from Shevchenko's \citep[see][]{1997SoSyR..31..219S} 
  magnitude-phase relationship. We assume that this applies to both
  Didymos surface and the 
  ejected particles. On the other hand, as already stated,
  the density of the particles is
  assumed to be the same as the bulk density of the asteroid, i.e., 
  $\rho_p$=2170  $\mathrm{kg\; m^{-3}}$.   

The ejected particles are
assumed to be distributed following a differential power-law size
distribution function, i.e., $n(r)\propto r^\kappa$, where we assume
an index $\kappa$=--3.5, and having a 
broad size range from $a_p$=1 $\mu$m to $a_p$=1 cm. Those 
values can be considered as typical for the dust grains ejected from active
asteroids \citep[see e.g.][and references
  therein]{2019A&A...624L..14M}.  The dust ejection is assumed to 
proceed instantly at the impact time, i.e., no particle emission  
is considered afterwards. 

The impact will generate a crater on Dimorphos surface, from which the
particles will be ejected. This is a primary ejection mechanism,
although more complex mechanisms surely will also be playing a 
role, such as impact-induced seismic waves
propagation across the body, which is able to  
produce lift-off of small particles due to shaking  
\citep{2022Tancredi}. In the event that this shaking mechanism 
is dominating the particle ejection, the expected speeds would be of
the order of Dimorphos escape speed, and the dust mass released would
be comparable to  
that produced by the impact itself \citep{2022Tancredi}.  While a large
range of ejection speeds might be expected, the fast moving material will
get dispersed very rapidly in space, being difficult to detect,
but the low-speed component of the ejecta will 
contribute to form a detectable dust plume, in a way similar to 
impacted active asteroids such as (596) Scheila
\citep{2011ApJ...733L...4J,2011ApJ...738..130M}. Given the large 
uncertainty in the relative contribution of those mechanisms
\citep[see, e.g.][]{2018Icar..312..128Y}, and the speed 
distribution of the ejecta,  we will assume different scenarios and
will compute the corresponding coma (by aperture photometry) and tail
brightness at 
different dates following the impact date in an attempt to provide some
insight into the detectability of the phenomenon as seen from Earth.

  An order-of-magnitude estimate of
  the ejected mass by the impact is provided by the scaling laws of
  Housen and Holsapple  \citep[see][their figure
    4]{2011Icar..211..856H}. For impacts into 
  basalt powder and dry quartz sand, for ejecta
  speeds of the order of the mean value of the escape speeds of the 
  binary system, ($v_{ej} \sim$0.25  $\mathrm{m\; s^{-1}}$), we get 
  $v_{ej}/v_i$=4$\times$10$^{-5}$. Then, the ratio of
  the mass ejected with speeds higher than $v_{ej}$, $M_{ej}$, to the
  impactor mass is  $M_{ej}/m_i \sim$ 9000, so that $M_{ej} \sim$
  5$\times$10$^6$ kg. This assumes that the impactor and the target
  have the same bulk density, so it must be considered as an
  order-of-magnitude 
  approximation. For comparison, the Deep Impact 
  projectile, having $m_i$=370 kg and $v_i$=10.3  $\mathrm{km\;
    s^{-1}}$ (and
  therefore similar momentum to DART impactor) produced a dust cloud around
  comet 9P/Tempel 1 of 10$^6$ 
  kg \citep{2005Sci...310..274S}, i.e., the same order
  of magnitude than assumed here. 
  
The ejection speed distribution is unknown. According to both modelling of
  impacted asteroids such as (596)
  Scheila \citep{2011ApJ...738..130M} and laboratory
  collision experiments \citep{1998P&SS...46..921G} there seems to be
  a weak dependence of particle speed with size. We will assume
  different ejection speeds starting from the escape velocity of
  Dimorphos ($v_{esc}$=0.09  $\mathrm{m\;
    s^{-1}}$). \cite{2001Icar..154..402K}, from 
  their impact experiments into ice-silicate surfaces, gave
  an upper limit to the ejecta speeds of the order of 700-800
  $\mathrm{m\; s^{-1}}$,
  independently of projectile speeds, that we adopt as an upper limit
  in our computations.

\section{Comparison with our Monte Carlo dust tail code}

\begin{figure*}
    \includegraphics[angle=-90,width=\textwidth]{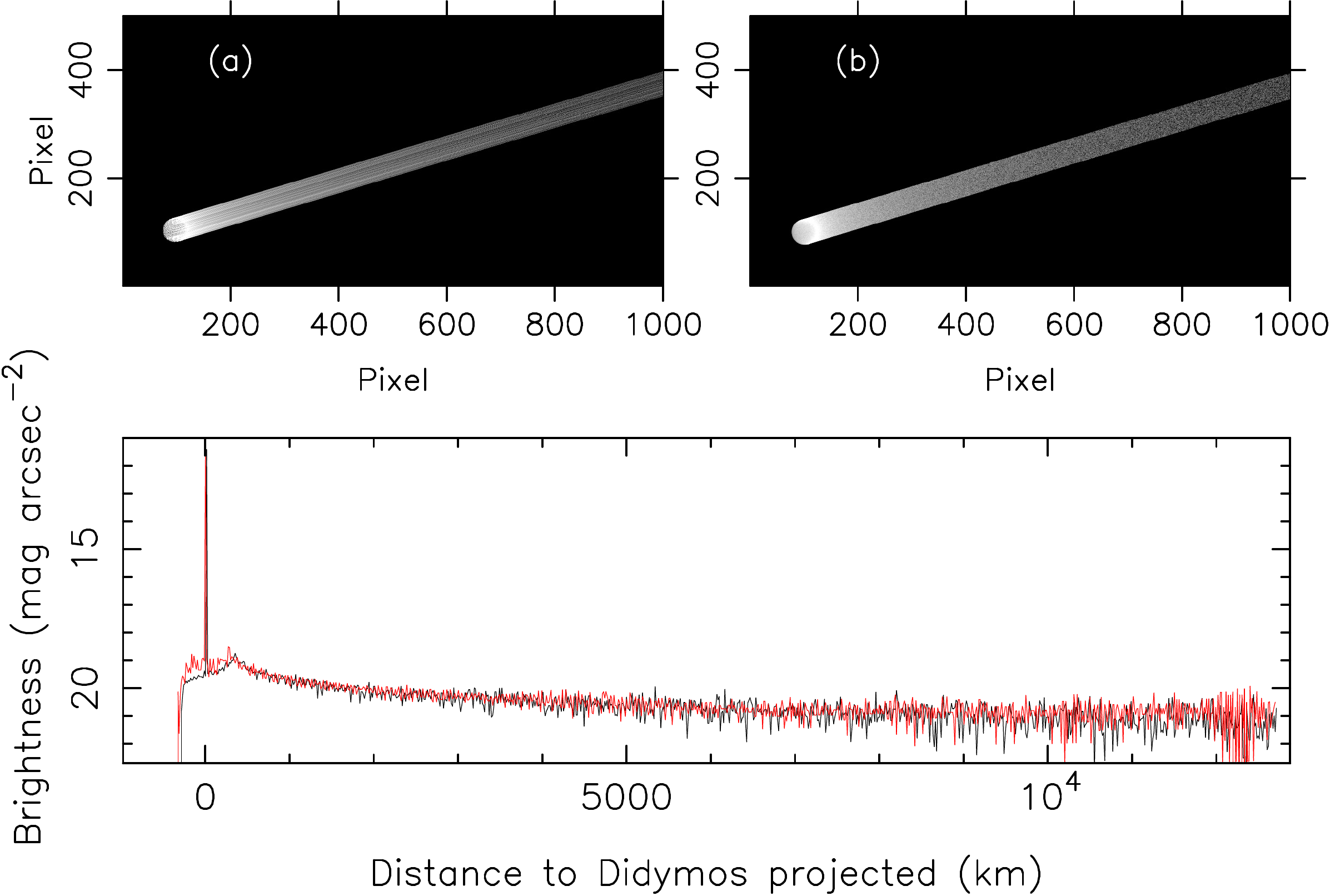}\\
    \caption{Simulated tail images of the Didymos system as would be
      seen from Earth after DART impact (Sept. 26.986, 2022) on
      October 2nd, 2022, with the detailed method consisting on the
      integration of equation ~\ref{eqtraj} (panel (a)), and the Monte
      Carlo dust tail code (panel (b)) considering only solar gravity
      and radiation pressure forces. The lowermost panel depicts a
      comparison between brightness scans along the tail for images
      (a) (red line) and (b) (black line), as a function of distance
      to Didymos projected. The spikes at zero projected distance is
      caused by the Didymos surface brightness. The orientation of the
      images is the 
      conventional: celestial North up, East to the left.}  
    \label{compcarlos}
\end{figure*}

 As another check of the calculations, our method was compared with
 the outcome of the Monte Carlo dust tail 
code that we routinely use to retrieve the dust environment from active
asteroids and comets 
\citep[see e.g.][and references 
    therein]{2021MNRAS.506.1733M}. In such a code, the only forces
  considered on particles are the solar gravity and radiation
  pressure, so that their trajectories around the Sun are Keplerian,
  and the numerical 
  simulation proceeds way faster than with the detailed
  trajectory calculations provided here with the integration of the
  equation of motion ~\ref{eqtraj}. For ejection
  speeds larger than the 
  escape velocity of the binary system the two
  methods should give similar results, since at such speeds only a
  small fraction of 
  the ejected mass is delivered to Didymos and Dimorphos, as we will
  see later (see Table ~\ref{massfate}), and the gravity forces of the
  small binary components is not very important. Figure ~\ref{compcarlos}
  displays a comparison between the two methods, for simulated tails 
  corresponding to a ejection speed of $v$=0.7  $\mathrm{m\; s^{-1}}$, for an
  observation date of October 2nd, 2022, with the rest of 
  physical parameters as stated above. Agreement is very good between
  the two methods. Therefore we can trust the new, more accurate,
  approach. For speeds near the escape
  velocity of the system or smaller, the comparison is meaningless, as
  the trajectory of particles would differ, and, besides, a considerable
  fraction of the ejected mass is actually transferred to the two binary
  components (see Table  ~\ref{massfate}). 

\section{Results and discussion}

\subsection{Dust tail simulations}

First of all, we generate the synthetic dust tail that, under
low-speed ejecta conditions, 
would be observable from the Earth.  Simulations are performed at
various observing dates, from one day 
after the impact up 
to five days later, at four different ejection speeds namely the Dimorphos
escape velocity, and twice, four, and eight times that value. Table 
~\ref{observinglog} shows the 
assumed circumstances for the three dates. The plate scale in
$\mathrm{km\; px^{-1}}$  of the last column has been determined
assuming a typical 
spatial resolution  
of 0.25 \arcsec px$^{-1}$. 


  \begin{table}
	\centering
	\caption{Circumstances of the observations assumed}
        \label{observinglog}
	\begin{tabular}{lllll} 
	  \hline
         Date  & r  & $\Delta$  & Phase  & Scale \\
          (UT) & (au) & (au) & angle ($^\circ$) &  ( $\mathrm{km\; px^{-1}}$) \\
         \hline
         Sep/28.0/2022 & 1.043 & 0.075 & 54.8 & 13.7 \\
         Sep/30.0/2022 & 1.039 & 0.073 & 57.8 & 13.4 \\
         Oct/02.0/2022 & 1.034 & 0.072 & 60.7 & 13.2 \\
		\hline
	\end{tabular}
\end{table}

  The model computes the asteroid tail brightness at a given date,
  where the asteroid nucleus (Didymos) brightness is also included in
  the simulated 
  images, using expression ~\ref{eqbright} with $a_p \equiv R_N$=390 m, the Didymos radius, and geometric albedo
  and linear phase coefficient as stated above for the particles. We
  neglect the 
  contribution of Dimorphos, since it is only a 4\% of
  the Didymos cross section.  

\begin{figure*}
    \includegraphics[angle=-90,width=\textwidth]{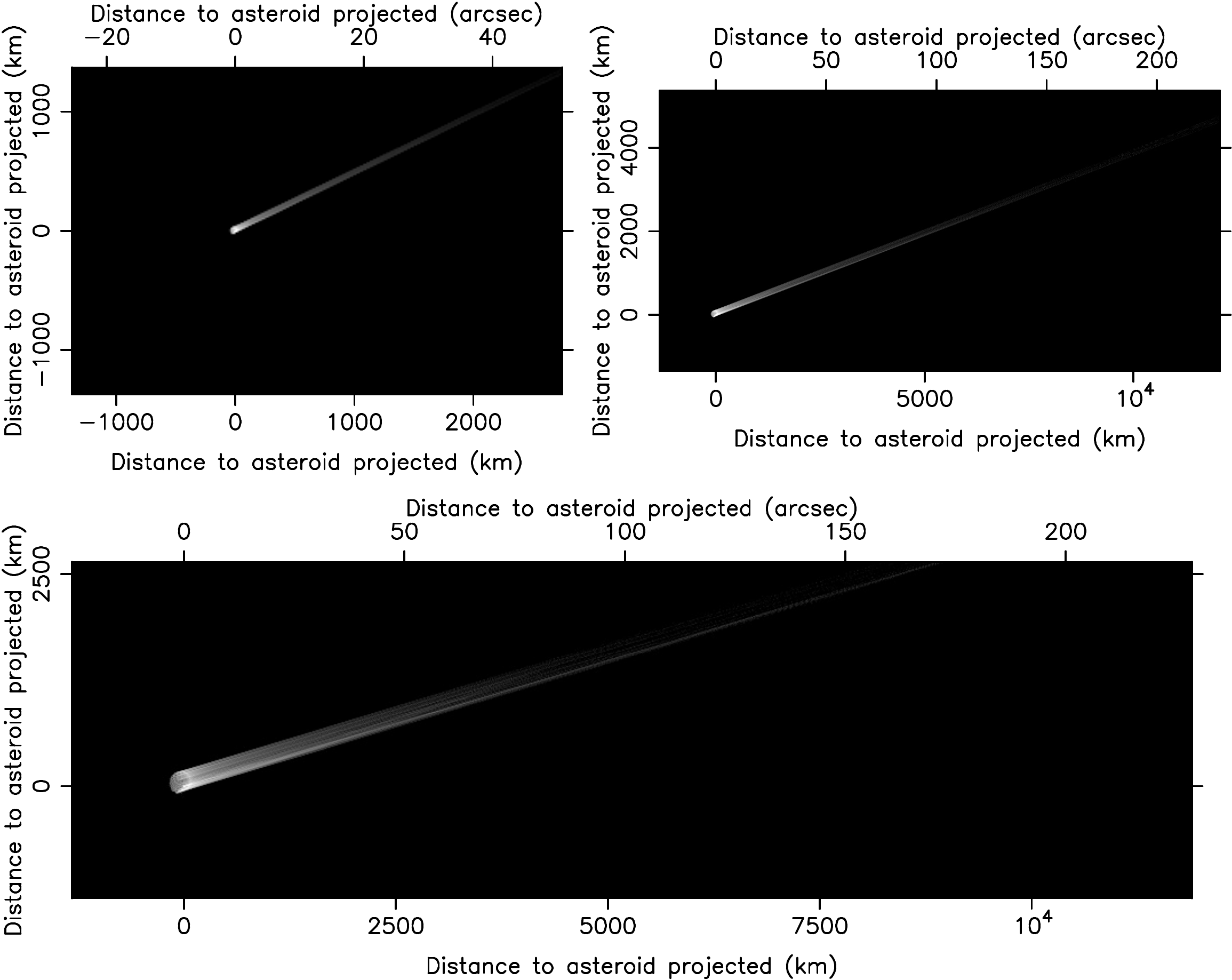}\\
    \caption{Simulated tail images of Didymos system as would be seen from Earth
      after DART impact (assumed to occur on 2022 Sept. 26, 23:14 UT), at
      three observing dates,  
      2022 Sept 28 (top left), Sept. 30 (top right), and Oct. 2
      (bottom), assuming an ejection speed from Dimorphos surface
      of $v_{ej}$=0.36  $\mathrm{m\; s^{-1}}$, corresponding to four
      times the escape 
      velocity of Dimorphos. The plate scale is assumed at 
      0.25 \arcsec px$^{-1}$, and the filter used is a
      $r$-Sloan. The corresponding brightness range are to 16-13, 19-16, and
      20-17 $\mathrm{mag\; arcsec^{-2}}$, respectively. The images are all
      oriented North up, East to the left.} 
    \label{threetails}
\end{figure*}

Figure ~\ref{threetails} depicts the appearance of the dust tails at the
three first 
observing dates in Table ~\ref{observinglog}, for an assumed ejection
speed of four times the Dimorphos escape velocity
(0.4 $\mathrm{m\; s^{-1}}$), which turns out to
be slightly higher than the escape speed of the binary system, and for the size
distribution, total mass ejected, and 
particle physical properties as given above. If those most favourable conditions
prevail, a narrow,  $\lesssim$1\arcmin ~in length, tail would be
detectable up to at least 5 days after 
impact, as the brightness levels would be well
below 20 $\mathrm{mag\; arcsec^{-2}}$. For
higher dust mass ejected, the tail detection would be obviously
easier.  

\subsection{Aperture photometry}

Aperture photometry also provides very important information on the
amount of material 
ejected and the fraction of such mass which is sent to free space,
or, depending on the ejecta velocity, is transferred to Didymos
or Dimorphos. We have calculated the variation of the predicted
magnitude, as a function of time since DART impact, for several ejecta
speeds ranging from the escape velocity of Dimorphos ($v_{esc}$=0.09
$\mathrm{m\; s^{-1}}$) to the upper
limit set by the collision experiments mentioned above
\citep{2001Icar..154..402K}, $v$=800  $\mathrm{m\; s^{-1}}$. In the low-speed
ejecta regime, i.e., speeds from the escape velocity of Dimorphos
up to a few times that value, and for the three observing dates shown
in Table ~\ref{observinglog}, we computed the $r$-Sloan magnitudes
using 1000 km radii ($\sim$18\arcsec) apertures. The results are shown in Table
~\ref{photometry1}, together with the unperturbed magnitude that would
be attained in the absence of the DART collision. We can see that
under low-speed ejecta conditions the ejecta cloud would give a rather
easily detectable signal, with about 
3 to 1.4 magnitudes below the ``clean'' system since the time of
impact to 5 days later, respectively. In the case of Deep Impact on comet
9P/Tempel 1, the object experienced a sudden increase in brightness of
2.5 magnitudes after the impact \citep{2005Sci...310..265M}, in
line with that found here. We underline that the low-speed ejecta 
regime would be prevailing if the shaking mechanism
\citep{2022Tancredi} is dominant.

  \begin{table}
	\centering
	\caption{Calculated $r$-Sloan magnitudes after DART impact,
          as a function of date
          and ejection speed, relative to Dimorphos escape
          velocity ($v_{esc}$=0.09  $\mathrm{m\; s^{-1}}$), compared to the
          unperturbed system magnitude.}
        \label{photometry1}
	\begin{tabular}{cccccc} 
	  \hline
      Date  & $v$=$v_{esc}$  & $v$=2$v_{esc}$  & $v$=4$v_{esc}$  &
      $v$=8$v_{esc}$  & Dust-free \\ 
          (UT) &  &  &  & &  mag \\
         \hline

         Sep/28.0/2022 & 11.66 & 11.36 & 11.28 & 11.23 & 14.30 \\
         Sep/30.0/2022 & 13.17 & 12.68 & 12.42 & 12.37 & 14.34 \\
         Oct/02.0/2022 & 13.96 & 13.22 & 12.97 & 12.92 & 14.39 \\
         \hline
	\end{tabular}
\end{table}

\begin{table}
	\centering
	\caption{Dust mass delivered to Didymos and Dimorphos as a
          function of date and ejection speeds.}
        \label{massfate}
	\begin{tabular}{cccc} 
	  \hline
      Date  & Mass to   &  Mass to  & Mass to   \\
          (UT) & Didymos  & Dimorphos  &  space \\
      \hline
      \multicolumn{4}{c}{$v$=$v_{esc}$}\\
      \hline      
  Sep/28.0/2022 & 7.9$\times$10$^5$ & 1.5$\times$10$^6$ & 2.7$\times$10$^6$ \\  
  Sep/30.0/2022 & 1.8$\times$10$^6$ & 1.6$\times$10$^6$ & 1.6$\times$10$^6$ \\  
  Oct/02.0/2022 & 2.4$\times$10$^6$ & 1.7$\times$10$^6$ & 9.0$\times$10$^5$ \\
  \hline
      \multicolumn{4}{c}{$v$=2$v_{esc}$} \\
      \hline      
  Sep/28.0/2022 & 1.4$\times$10$^6$ & 8.9$\times$10$^3$ & 3.6$\times$10$^6$ \\  
  Sep/30.0/2022 & 1.7$\times$10$^6$ & 1.7$\times$10$^4$ & 3.3$\times$10$^6$ \\  
  Oct/02.0/2022 & 1.9$\times$10$^6$ & 2.0$\times$10$^4$ & 3.1$\times$10$^6$ \\  
  \hline
      \multicolumn{4}{c}{$v$=4$v_{esc}$} \\
      \hline      
  Sep/28.0/2022 & 3.7$\times$10$^5$ & 4.6$\times$10$^4$ & 4.6$\times$10$^6$ \\  
  Sep/30.0/2022 & 4.9$\times$10$^5$ & 4.9$\times$10$^4$ & 4.5$\times$10$^6$ \\  
  Oct/02.0/2022 & 5.4$\times$10$^5$ & 5.3$\times$10$^4$ & 4.4$\times$10$^6$ \\  
  \hline
      \multicolumn{4}{c}{$v$=8$v_{esc}$}\\
      \hline      
  Sep/28.0/2022 & 1.5$\times$10$^5$ & 1.4$\times$10$^1$ & 4.9$\times$10$^6$ \\  
  Sep/30.0/2022 & 1.5$\times$10$^5$ & 1.4$\times$10$^1$ & 4.9$\times$10$^6$ \\  
  Oct/02.0/2022 & 1.5$\times$10$^5$ & 1.4$\times$10$^1$ & 4.9$\times$10$^6$ \\  
      \hline      
	\end{tabular}
\end{table}

\begin{figure*}
    \includegraphics[angle=-90,width=0.90\textwidth]{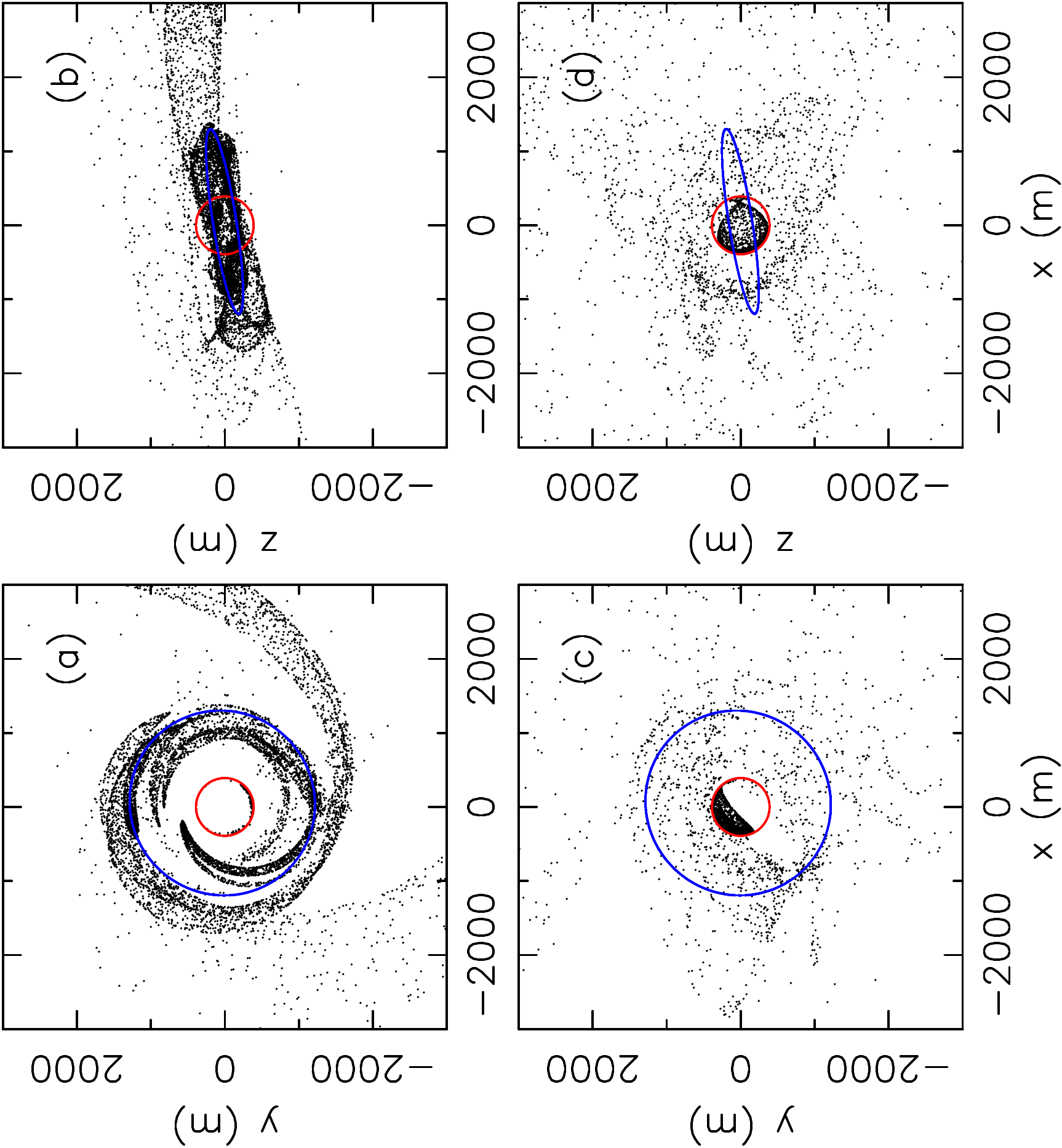}\\
    \caption{Projection on the $x-y$ and $x-z$
      planes of the final position of $r$=1 cm particles on September 28th,
      2022. The red circle represents Didymos, and the blue ellipse 
      represents the orbital path of Dimorphos.  
      The upper panels (a) and (b) correspond to ejection velocity
      $v$=$v_{esc}$, and the lower panels (c) and (d) to
      $v$=$2v_{esc}$, where $v_{esc}$=0.09  $\mathrm{m\; s^{-1}}$, the escape
      velocity of Dimorphos.} 
    \label{traj1v2v}
\end{figure*}

\begin{figure}
    \includegraphics[angle=-90,width=0.5\textwidth]{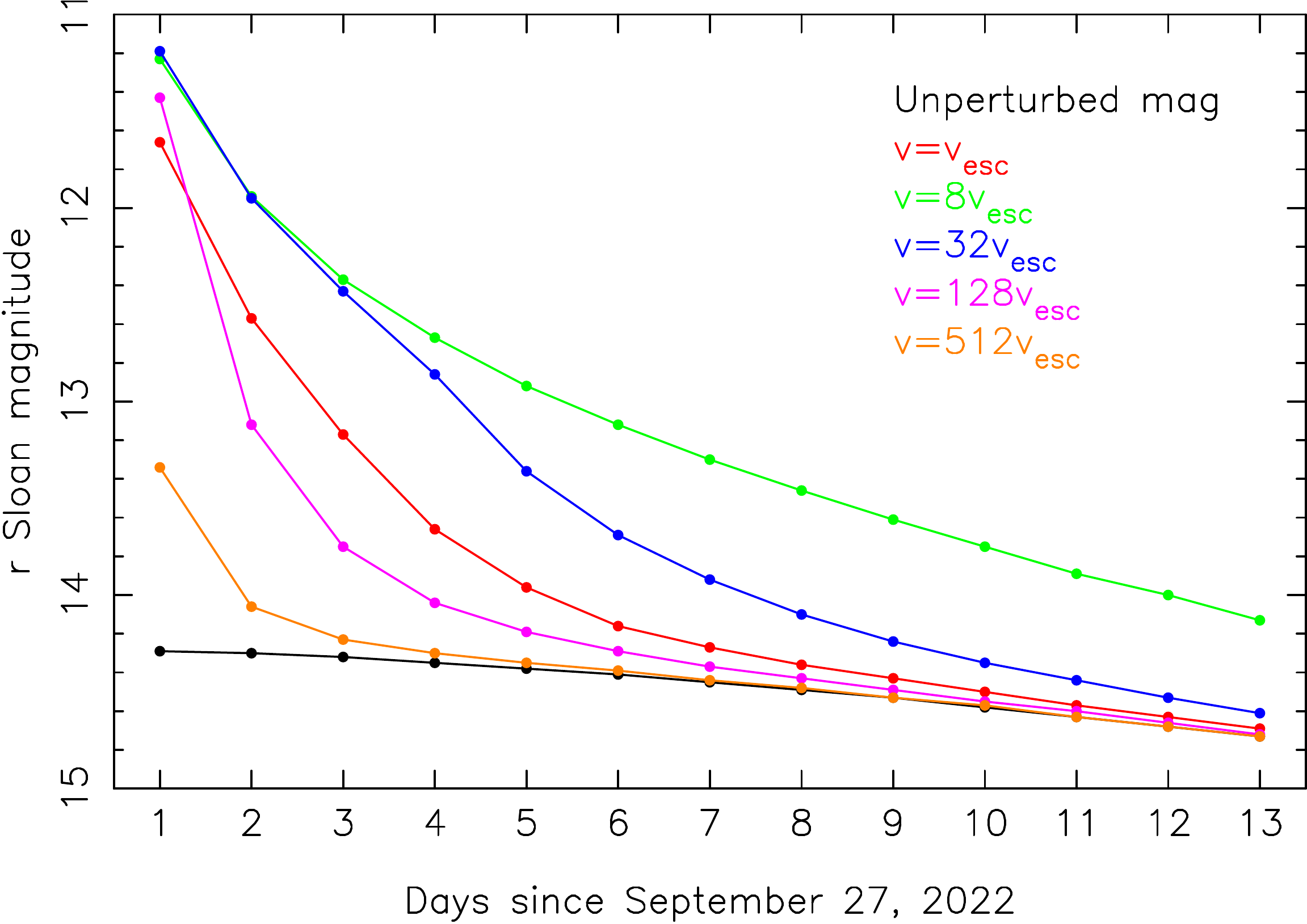}\\
    \caption{Calculated $r$-Sloan magnitudes as a function of date and
    ejection velocities, for relatively slow ejection speeds, compared
    to the unperturbed system magnitude.}  
    \label{magevol_slow}
\end{figure}

\begin{figure}
    \includegraphics[angle=-90,width=0.5\textwidth]{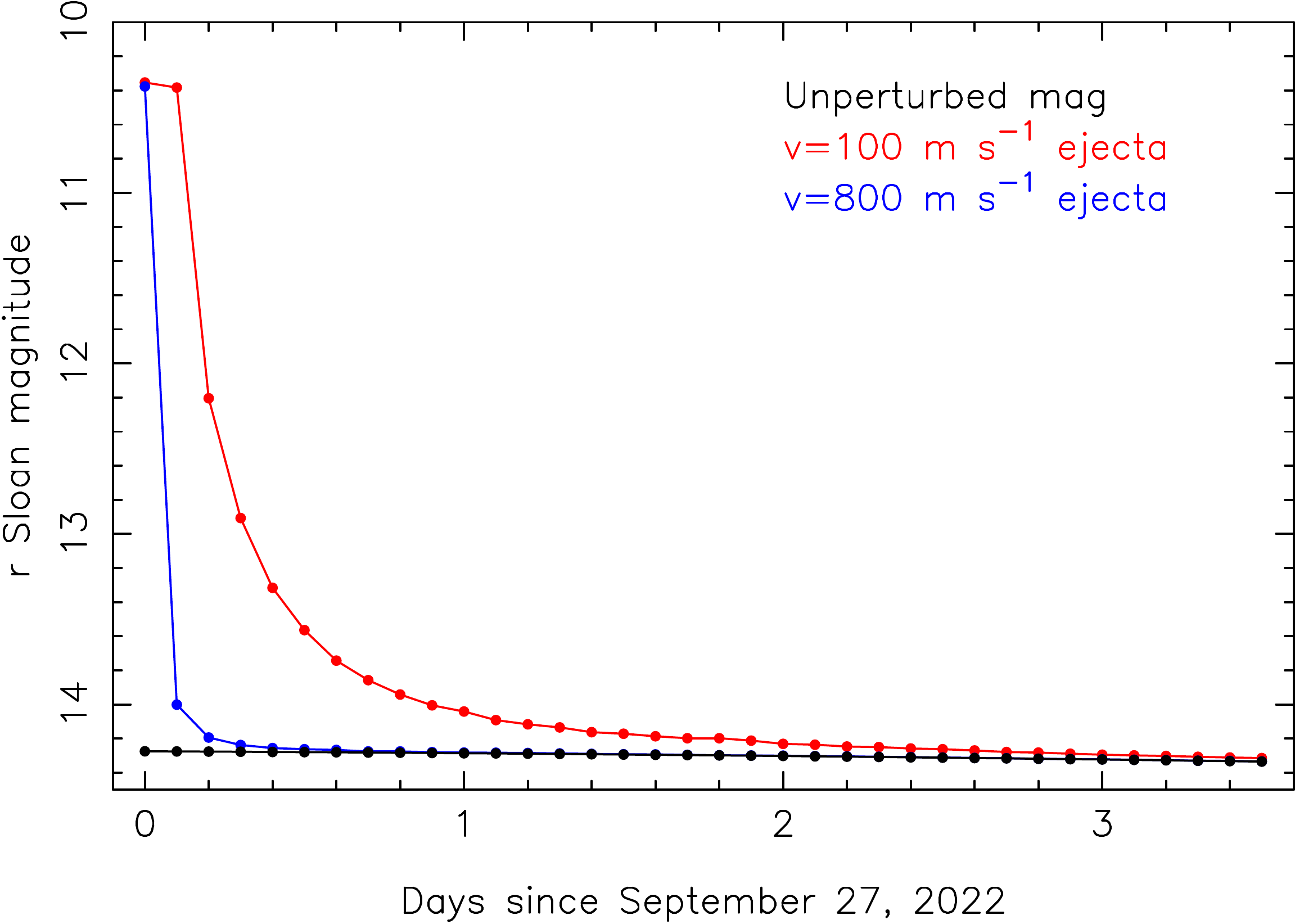}\\
    \caption{Calculated $r$-Sloan magnitudes as a function of date and
      ejection velocities, for large ejection speeds,
      compared to the unperturbed system magnitude.} 
    \label{magevol_fast}
\end{figure}

\begin{figure}
    \includegraphics[angle=-90,width=0.5\textwidth]{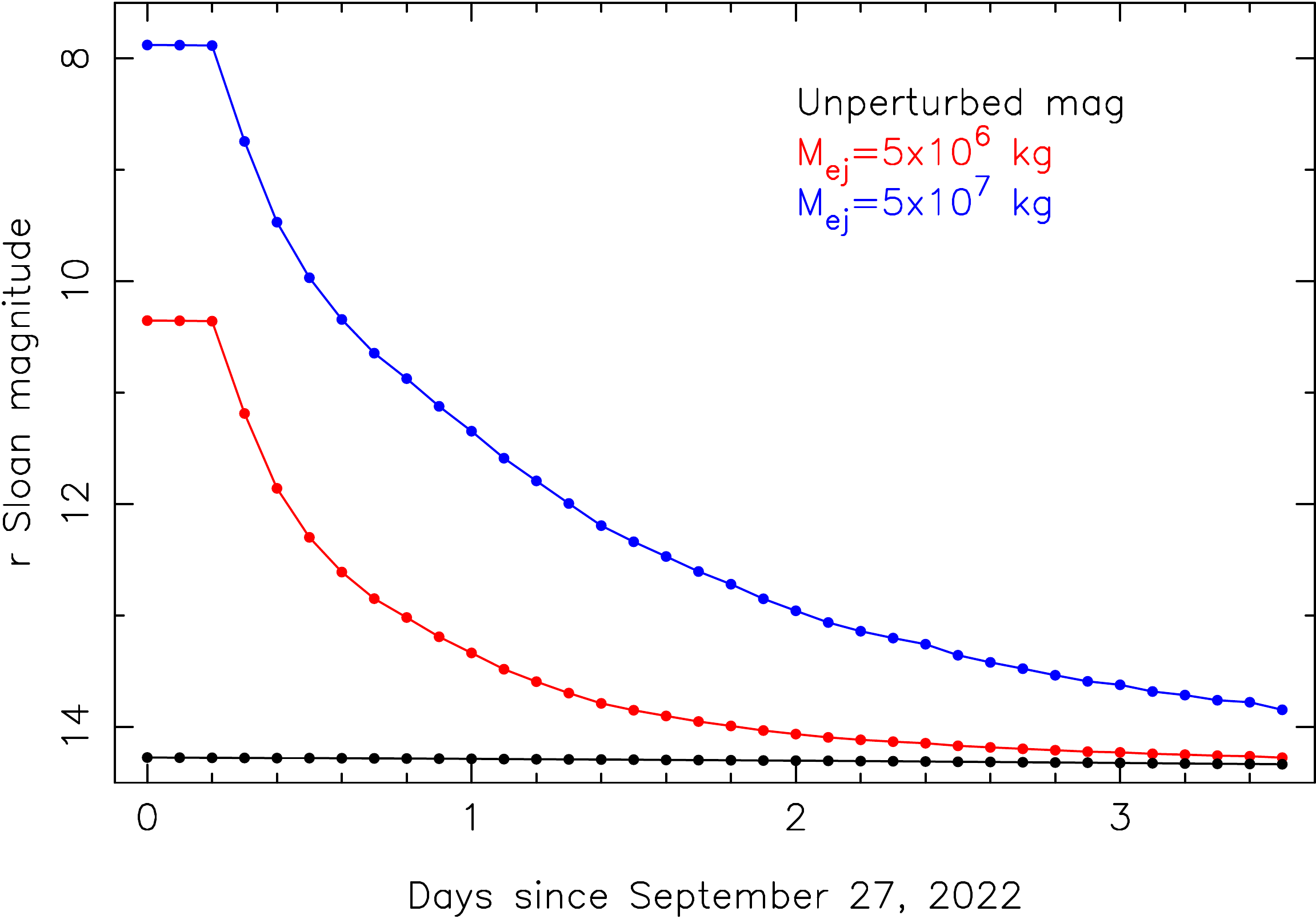}\\
    \caption{Calculated $r$-Sloan magnitudes as a function of date and
      total ejected masses of $M_{ej}$=5$\times$10$^6$ kg and
      $M_{ej}$=5$\times$10$^7$ kg, 
      for an ejection speed of $v$=512$v_{esc}$=45  $\mathrm{m\; s^{-1}}$,    
      compared to the unperturbed system magnitude.} 
    \label{magevol_2mass}
\end{figure}

It is
interesting to see that the highest brightness level does not occur at
the minimum velocity assumed, $v$=$v_{esc}$,
but corresponding to four to eight times that value. This can be
explained by the fact that 
a significant fraction of the mass ejected at such speeds is
transferred either to Didymos or Dimorphos. Thus, in Table
~\ref{massfate}, where we show the fate of the ejected mass, we see
that a significant fraction of the ejected mass, up to $\sim$50\% in
some cases, is sent to one of the binary components when the ejecta is
released at speeds of $v$=$v_{esc}$ or $v$=$2v_{esc}$. For larger
speeds, we wee that most of the mass is sent to space. In addition, it
is also interesting to note that the fate of the mass is different for
$v$=$v_{esc}$ and for $v$=$2v_{esc}$. In the former case, a
considerable fraction of the mass is, depending on the date,  either
transferred to Dimorphos (on Sept. 28), or it 
is shared between Didymos and Dimorphos (Sept. 30, and Oct. 2), while
in the latter case, a considerable amount of mass is sent to Didymos
while a negligible fraction lands on Dimorphos at all the observing
dates. This can be clearly 
seen in Figure ~\ref{traj1v2v} for the first observing date, Sept. 28. In the 
upper panels (a) and (b), the projections on the $x-y$ and $x-z$
planes of the positions of the particle at the end of integration
for $v$=$v_{esc}$, are given. We see that only
a few particles stick to Didymos (red circle), while many of them are
near the orbital path of Dimorphos, eventually colliding with it. On
the contrary, in 
the lower panels (c) and (d), that correspond to $v$=$2v_{esc}$, we see that
many particles got sticked to Didymos, while there are only a few
particles near Dimorphos orbit.

At speeds higher than $v$=$8v_{esc}$, the brightness decreases 
as a function of date at a faster rate, because the particles tend to leave
the aperture limits. Figure ~\ref{magevol_slow} gives the
lightcurves from ejection speeds of $v$=$v_{esc}$ up to
$v$=$512v_{esc}$=45  $\mathrm{m\; s^{-1}}$, along with the magnitude of the
unperturbed system, using 1000 km aperture radius in all cases. In the
most favourable case of $v\sim 8v_{esc}$, the 
ejecta would give a detectable signal during 13 days or more after impact,
but for the higher speed cases 
the event would be detectable only during 1 day (for $v=512 v_{esc}$)  to 10 days
(for $v=32 v_{esc}$) since impact time.  

At still higher speeds than those corresponding to  Figure
~\ref{magevol_slow}, the decrease of brightness is dramatic in only a
few hours after impact. In Figure ~\ref{magevol_fast}, lightcurves for
ejecta speeds of 100 and 800  $\mathrm{m\; s^{-1}}$, as a function of time, are
shown. For $v$=100  $\mathrm{m\; s^{-1}}$, the ejecta would be detectable within 
one day after impact, while in the least favourable scenario, $v$=800
$\mathrm{m\; s^{-1}}$, the observing window reduces to only $\sim$5
hours at most. 

When the total ejected mass is different to that estimated from  
scaling laws ($M_{ej} \sim$5$\times$10$^6$ kg), the observed
magnitudes would obviously differ from those shown in the previous
graphs. To have an estimate on how much the evolution could be different,
in Figure ~\ref{magevol_2mass} we display the results for an
intermediate ejection speed of  $v$=$512v_{esc}$=45  $\mathrm{m\; s^{-1}}$, for
the nominal mass and for ten times higher that value ($M_{ej}
\sim$5$\times$10$^7$ 
kg). In the latter case, a decrease of $\sim$2.5 mag with respect to
the nominal mass is foreseen. The flux increment then becomes less and
less pronounced as the time passes, the difference between both cases
being of only 0.4 magnitudes after 3.5 days. Similar differences are
expected for other ejection speeds. Therefore the increase in
brightness and its evolution with time will be an useful indicator of
the amount of material ejected and the distribution of ejection speeds.

\section{Conclusions}

Simulations of the dust cloud generated after DART impact on Dimorphos
satellite of (65803) Didymos have been performed. The calculations
have been made by integrating the equation of motion of a large
amount of individual particles ejected from Dimorphos. The total
ejected dust mass is 
calculated on the basis of known scaling laws, being about
5$\times$10$^6$ kg. A power-law size distribution function of power
index --3.5 and limiting radii of 1 $\mu$m and 1 cm is assumed. Depending on the
ejection speed, we give insight on the detectability of such  
cloud for Earth-based observers.  The most favourable ejection
scenario for longer term 
detection would occur when a considerable fraction of the total mass
is ejected at speeds of the order of a few times the escape
speed of Dimorphos. In such case, aperture photometry measurements
would reveal an observed brightness well above the
unperturbed system brightness by 1-3 magnitudes for up to 5 days after
impact, and a narrow dust tail of length $\lesssim$1\arcmin
 ~with 20 $\mathrm{mag\; arcsec^{-2}}$ minimum brightness would be
seen in the sky. That would be the case if most of the ejecta is caused by
impact-induced seismic waves, producing seismic shaking of the
satellite, as the predicted speeds would be below 1 $\mathrm{m\; s^{-1}}$. In
the case that the ejecta travels at much larger speeds, of the
order of 100  $\mathrm{m\; s^{-1}}$, 
the brightness increase will be detected only within 2 days after impact
time, inducing a fast-decreasing lightcurve from almost 4
to 0.3 magnitudes above the baseline brightness in that time frame. For the
highest speeds attainable (following impact experiments) of 800
$\mathrm{m\; s^{-1}}$, the observing window reduces to only $\sim$5
hours after impact at
most. This model  
will serve to constrain the ejecta properties when the corresponding
flux measurements become available. The monitoring of the tail and
coma brightness in the hours and early days after impact will be
relevant for understanding the small particle ejection mechanisms and
its contribution to the impact process.  

\section*{Acknowledgements}

We are indebted to the anonymous referee for his/her constructive
comments that help to improve the manuscript. 

FM acknowledges financial support from the State Agency for
Research of 
the Spanish MCIU through the "Center of Excellence Severo Ochoa" award
to the Instituto de Astrof\'\i sica de Andaluc\'\i a (SEV-2017-0709).
FM also acknowledges financial support from the  Spanish  Plan
Nacional  de  Astronom\'\i a  y  Astrof\'\i sica  LEONIDAS
project RTI2018-095330-B-100,  and  project  P18-RT-1854  from  Junta
de  Andaluc\'\i a. ACB and PYL acknowledge funding by the
SU-SPACE-23-SEC-2019 EC-H2020 NEO-MAPP project (GA 870377).  ACB also
acknowledges funding by the Spanish Ministerio de Ciencia e Innovaci\'on
RTI2018-099464-B-I00 project. GT and BD acknowledge finantial support
from project FCE-1-2019-1-156451 of the Agencia Nacional de
Investigaci\'on e Innovaci\'on ANII (Uruguay). 

This  work  has  made  use  of  NASA's  Astrophysics  Data System
Bibliographic Services and of the JPL's Horizons system.

\section{Data availability}
This work uses simulated data, generated as detailed in the text.


\bibliographystyle{mnras}
\bibliography{segundo-revised}

\appendix

\section{Dust tail brightness computation}

We start by defining 
    a high-resolution grid of particle radius bins logarithmically
    distributed in 
    the interval [$r_{min}$,$r_{max}$], where $r_{min}$ and $r_{max}$ are
    the limits of the particle size distribution function assumed. For each
    radius bin $[r_1,r_2]$, we computed the trajectories of sampled particles
    having a mean radius in such bin, $r_{mean}$, whose dynamical
    behaviour is assumed to be 
    representative of the full $[r_1,r_2]$ range. The number of radius
    bins was set to 4000. On Dimorphos surface we define a 
    latitude-longitude grid of 
    similar area cells. The particles are ejected radially away from
    the centre of each cell. The number of latitude$\times$longitude 
    surface elements was set to 20$\times$20.
    This isotropic ejection pattern can be set to a more 
    realistic ejection pattern once the  
exact geometry of the impact is known, probably very close to the
impact time. 

The flux contribution (expressed in
magnitudes, $m$) of a spherical particle of radius $a_p$ (in
metres) falling on a certain pixel of the image on the $(N,M)$
photographic plane is given by the following expression:

\begin{equation}
 p_R\pi a_p^2=\frac{2.24 \times 10^{22} \pi r_h^2\Delta^2 10^{0.4
     (m_\odot-m)}}
     {G(\alpha)}
     \label{eqbright}
\end{equation}

\noindent
where $r_h$ is the asteroid heliocentric distance, $\Delta$ is the
asteroid geocentric distance (both in au), and $m_\odot$ is the apparent solar
magnitude in the filter used. All the simulated images in this work
refer to $r$-Sloan magnitudes, for which  $m_\odot$=--26.95
\citep{2001AJ....122.2749I}.  The particle geometric albedo at zero
phase angle is given by $p_R$, and $G(\alpha)=10^{-0.4\alpha\phi}$ is
the phase angle correction, where $\alpha$ is the phase angle, and
$\phi$ is the linear phase coefficient. The contribution to the
brightness, $B$, in the mentioned pixel, in units of mag arcsec$^{-2}$
would be $B=m+2.5\log_{10}A$, 
where $A$ is the projected pixel area on the sky in arcsec$^2$. To
perform the calculations within the computer code, we find most
appropriate to work on units of the mean solar
disk intensity units in the corresponding wavelength range,
$i/i_\odot$, where 
$i_\odot=F_\odot/\pi$, being $F_\odot$ the radiation flux density
at the solar surface. Knowing that the solid angle of the Sun disk at
a distance of 1 au is $S_a$=2.893$\times$10$^{6}$ arcsec$^2$, the relation between $i/i_\odot$ and $S$ becomes \citep[e.g.][]{1992A&A...260..455J}:

\begin{equation}
 S=C + m_\odot -2.5 \log_{10} i/i_\odot
\label{eqJockers}
\end{equation}

\noindent
where $C=2.5\log_{10} S_a=16.153$.


The mass ejected from each surface cell on Dimorphos is given by
$M_{cell}=M_{ej} S_{cell} (4\pi R_{sec}^2)^{-1}$, where $M_{ej}$ is
the total dust mass ejected and $S_{cell}$ is
the cell surface area. The number of particles ejected from each
cell in the bin radius $[r_1,r_2]$ will be given by:

\begin{equation}
N[r_1,r_2]= \frac{3 M_{cell}}{4 \pi \rho_p} \frac{\displaystyle \int_{r_1}^{r_2} r^\kappa dr}{\displaystyle \int_{r_{min}}^{r_{max}} r^{\kappa+3} dr}
\end{equation}
where $\kappa$ is the index of the assumed power-law differential size
distribution function. Then, 
if at the end of its trajectory, a particle of radius $r_{mean}$ falls
within a certain pixel, the brightness on that
pixel would be incremented by the product of $i/i_0$ (equation ~\ref{eqJockers}) and
$N[r_1,r_2]$. The procedure is then carried out for all the 4000 particles
ejected from each of the 20$\times$20 aforementioned surface cells on Dimorphos.

\bsp	
\label{lastpage}

\end{document}